\title{Construction of PMDS and SD Codes extending RAID 5}
\author{Mario Blaum\\
IBM Almaden Research Center\\
San Jose, CA 95120
}
\documentclass[12pt]{article}
\usepackage{latexsym}
\usepackage{multirow}
\usepackage{color}
 \newtheorem{theo}{Theorem}[section]

 \newtheorem{defin}{Definition}[section]
 \newtheorem{ex}{Example}[section]

\newtheorem{COROLLARY}{\indent Corollary}

\newtheorem{EXAMPLE}{\indent Example}

\newtheorem{THEOREM}{\indent Theorem}

\newtheorem{REMARK}{\indent Remark}

\newcommand{\fullstop}{\hspace{-0.85em} {\bf .}}

\newcommand{\ue}{\mbox{$\underline{e}$}}

\newcommand{\uc}{\mbox{$\underline{c}$}}

\newcommand{\la}{\mbox{$\leftarrow$}}

\newcommand{\al}{\mbox{$\alpha$}}

\newcommand{\eq}{\mbox{$\, =\,$}}

\newcommand{\qed}{\hfill$\Box$\\[1ex]}
\newcommand{\pf}{{\bf Proof: }}

\newcommand{\xor}{\mbox{$\,\oplus\,$}}

\newcommand{\C}{\mbox{${\cal C}$}}

\newcommand{\cO}{\mbox{${\cal O}$}}

\newcommand{\br}{\\ }
\newcommand{\ce}{\begin{center}}
\newcommand{\cen}{\end{center}}

\newcommand{\ipb}{\begin{description}}
\newcommand{\ipn}{\end{description}}

\newcommand{\qb}{\begin{quote}}
\newcommand{\qn}{\end{quote}}

\newcommand{\tp}{\begin{titlepage}}
\newcommand{\tpn}{\end{titlepage}}
\newcommand{\zb}{\begin{figure}[hbtp]}
\newcommand{\zn}{\end{figure}}

\newcommand{\EQX}[1]{\begin{equation}\label{#1}}
\newcommand{\ENX}{\end{equation}}
\newcommand{\EQL}{\begin{eqnarray*}}
\newcommand{\ENL}{\end{eqnarray*}}
\newcommand{\EQLX}[1]{\begin{eqnarray}\label{#1}}
\newcommand{\ENLX}{\end{eqnarray}}

\newcommand{\open}{\begin{document}}
\newcommand{\close}{\end{document}}
\newcommand{\lfcr}[1]{\br\hspace*{#1em}}
\newenvironment{mat}[1]
{\left[\begin{array}{#1}}{\end{array}\right]}
\newcommand{\GAMMA}{\Gamma}
\newcommand{\DELTA}{\Delta}
\newcommand{\THETA}{\Theta}
\newcommand{\LAMBDA}{\Lambda}
\newcommand{\XI}{\Xi}
\newcommand{\PI}{\Pi}
\newcommand{\SIGMA}{\Sigma}
\newcommand{\UPSILON}{\Upsilon}
\newcommand{\PHI}{\Phi}
\newcommand{\PSI}{\Psi}
\newcommand{\OMEGA}{\Omega}
\newcommand{\bldgreek}[1]{\mbox{\boldmath $#1$}}
\newcommand{\bldbeta}{\bldgreek{\beta}}
\newcommand{\bldgamma}{\bldgreek{\gamma}}
\newcommand{\blddelta}{\bldgreek{\delta}}
\newcommand{\bldepsilon}{\bldgreek{\epsilon}}
\newcommand{\bldvarepsilon}{\bldgreek{\varepsilon}}
\newcommand{\bldzeta}{\bldgreek{\zeta}}
\newcommand{\bldeta}{\bldgreek{\eta}}
\newcommand{\bldtheta}{\bldgreek{\theta}}
\newcommand{\bldvartheta}{\bldgreek{\vartheta}}
\newcommand{\bldiota}{\bldgreek{\iota}}
\newcommand{\bldkappa}{\bldgreek{\kappa}}
\newcommand{\bldlambda}{\bldgreek{\lambda}}
\newcommand{\bldmu}{\bldgreek{\mu}}
\newcommand{\bldnu}{\bldgreek{\nu}}
\newcommand{\bldxi}{\bldgreek{\xi}}
\newcommand{\bldpi}{\bldgreek{\pi}}
\newcommand{\bldvarpi}{\bldgreek{\varpi}}
\newcommand{\bldrho}{\bldgreek{\rho}}
\newcommand{\bldvarrho}{\bldgreek{\varrho}}
\newcommand{\bldsigma}{\bldgreek{\sigma}}
\newcommand{\bldvarsigma}{\bldgreek{\varsigma}}
\newcommand{\bldtau}{\bldgreek{\tau}}
\newcommand{\bldupsilon}{\bldgreek{\upsilon}}
\newcommand{\bldphi}{\bldgreek{\phi}}
\newcommand{\bldvarphi}{\bldgreek{\varphi}}
\newcommand{\bldchi}{\bldgreek{\chi}}
\newcommand{\bldpsi}{\bldgreek{\psi}}
\newcommand{\bldomega}{\bldgreek{\omega}}
\begin{document}
\parindent=10pt
\maketitle
\begin{abstract}
A construction of Partial Maximum Distance Separable (PMDS) and
Sector-Disk (SD) codes extending RAID 5 with two extra parities is
given, solving an open problem. Previous constructions relied on
computer searches, while our constructions provide a
theoretical solution to the problem.
\vspace{.3cm}

\end{abstract}

\section{Introduction}
\label{Introduction}
Consider an $m\times n$ array whose entries are elements in a finite field
$GF(2^b)$~\cite{ms} (in general, we could consider a field $GF(p^b)$, $p$ a
prime number, but for simplicity, we constrain ourselves to binary
fields). The $n$
columns represent storage devices like SSDs, HDDs
or tapes. The arrays (often called stripes also) are repeated as many
times as necessary. In order to protect against a device failure, a
RAID 4 or RAID 5 type of
scheme, in which one of the devices is the XOR of the other ones, can
be implemented. During reconstruction, the failed device is recovered
sector by sector. The problem with RAID 5 is, if an additional sector
is defective in
addition to the one corresponding to the failed device, data loss
will occur. A solution to this problem is using a second device for
parity (RAID 6), allowing for recovery against two failed devices.
However, this scheme may be
wasteful, and moreover, it is unable to correct the situation in
which in addition to the sector corresponding to the failed disk, we
have two extra failed sectors in the row (we always assume that
failed sectors can be identified, either by CRC or by other means, so
the correcting scheme is an erasure correcting scheme). In order to
overcome this problem,
the so called Partial MDS (PMDS) codes~\cite{bhh} and Sector-Disk
(SD) codes~\cite{pbh} were created. Very similar codes were presented
in~\cite{hsx}.

We start by giving the definition of PMDS and SD codes.

\begin{defin}
\label{defPMDS}
{\em
Let $\C$ be a linear $[mn,m(n-r)-s]$ code over a field such
that when codewords are taken row-wise as
$m\times n$ arrays, each row belongs in an
$[n,n-r,r+1]$ MDS code. Then,

\begin{enumerate}
\item
$\C$ is an $(r;s)$ partial-MDS (PMDS) code if,
{\em for any} $(s_1,s_2,\ldots,s_t)$ such that each $s_j\geq 1$
and $\sum_{j=1}^ts_j\eq s$, and for any $i_1,i_2,\ldots,i_t$ such that $0\leq
i_1<i_2<\cdots <i_t\leq m-1$, $\C$ can correct up
to $s_j+r$ erasures in each row $i_j$, $1\leq j\leq t$, of an array in $\C$.
\item
$\C$ is an $(r;s)$ sector-disk (SD) code if, for any
$l_1,l_2,\ldots,l_r$ such that
$0\leq l_1<l_2<\cdots <l_{r}\leq n-1$,
for any $(s_1,s_2,\ldots,s_t)$ such that each $s_j\geq 1$
and $\sum_{j=1}^ts_j\eq s$, and for any $i_1,i_2,\ldots,i_t$ such that $0\leq
i_1<i_2<\cdots <i_t\leq m-1$, $\C$ can correct up
to $s_j+r$ erasures in each row $i_j$, $1\leq j\leq t$, of an array in
$\C$ provided
that locations $l_1,l_2,\ldots l_{r}$ in each of the rows $i_j$
have been erased.
\end{enumerate}
}
\end{defin}

SD codes satisfy a weaker condition than PMDS codes, but they may be
sufficient in most applications. The case of $(r;1)$ PMDS codes has
been solved in~\cite{bhh}. In this paper, we address the case of
(1;2) PMDS and SD codes. Figure~\ref{figPMDS} illustrates the
difference between (1;2) PMDS and SD codes for a $4\times 5$ array
(i.e., a code of length 20): the array in the left depicts a
situation that can be handled by a (1;2) PMDS but not by a (1;2) SD code; the
second and the fourth rows have two erasures (denoted by $E$) each
and there is no column containing two of these erasures. The array in
the middle illustrates a situation in which the second and fourth
rows have two erasures each, but the second column contains two of
those erasures, which correspond to a total failure of the second
device. Individual erasures in a row can always be handled by single
parity (like in the first and the third rows). This situation can be
handled by both (1;2) PMDS and SD codes. Finally, the array in the right
shows the situation of three erasures in a row, and at most one in
the remaining ones. This situation can also be handled by both (1;2) PMDS
and SD codes (but not by RAID 6).

\begin{figure}
$$
\begin{array}{ccc}
\begin{array}{|c|c|c|c|c|}
\hline
1&0&1&0&0\\
\hline
E&1&E&0&1\\
\hline
1&1&1&0&1\\
\hline
1&E&1&1&E\\
\hline
\end{array}
&
\begin{array}{|c|c|c|c|c|}
\hline
1&E&1&0&0\\
\hline
0&E&1&0&E\\
\hline
1&E&1&0&1\\
\hline
E&E&1&1&1\\
\hline
\end{array}
&
\begin{array}{|c|c|c|c|c|}
\hline
1&0&1&E&0\\
\hline
E&1&E&E&1\\
\hline
1&1&1&E&1\\
\hline
1&0&1&E&1\\
\hline
\end{array}\\
\end{array}
$$
\caption{A $4\times 5$ array with different types of failures}
\label{figPMDS}
\end{figure}

In the next section we give the construction of both (1;2) PMDS and
SD codes. From now on, when we say PMDS or SD codes, we refer to
(1;2) PMDS or SD codes.

\section{Code Construction}
\label{construction}

Consider the field $GF(2^b)$ and let
$\al$ be an element in $GF(2^b)$.
The (multiplicative) order
of $\al$, denoted $\cO(\al)$, is the minimum $\ell$, $0<\ell$, such that
$\al^{\ell}\eq 1$. If $\al$ is a primitive element~\cite{ms}, then
$\cO(\al)\eq 2^b-1$. To each element $\al\in GF(2^b)$, there is an
associated (irreducible) minimal polynomial~\cite{ms} that we denote $f_{\al}(x)$.

Let $\al\in GF(2^b)$ and $mn\leq \cO(\al)$. Consider the $(m+2)\times mn$ parity-check matrix

\begin{eqnarray}
\label{pcSD}
\left(
\begin{array}{cccc|cccc|c|cccc}
\uc_0&\uc_1&\ldots&\uc_{n-1}&\uc_{n}&\uc_{n+1}&\ldots&\uc_{2n-1}&\ldots
&\uc_{(m-1)n}&\uc_{(m-1)n+1}&\ldots&\uc_{mn-1}\\
\end{array}
\right)
\end{eqnarray}
where $\uc_i$ denotes a column of length $m+2$, and, if $\ue_i$
denotes an $m\times 1$ vector whose coordinates are zero except for
coordinate $i$, which is 1, then, for $0\leq i\leq m-1$,

\begin{eqnarray}
\label{pcSDc}
\uc_{in},\uc_{in+1},\ldots ,\uc_{(i+1)n-1}&=&\left(
\begin{array}{cccccc}
\ue_i &\ue_i &\ldots &\ue_i & \ldots &\ue_i \\
\al^{in}&\al^{in+1}&\ldots &\al^{in+j}&\ldots &\al^{(i+1)n-1}\\
\al^{2in}&\al^{2in-1}&\ldots &\al^{2in-j}&\ldots &\al^{(2i-1)n+1}\\
\end{array}
\right)
\end{eqnarray}

We denote  as $\C^{(0)}(m,n;f_{\al}(x))$ the $[mn,m(n-1)-2]$ code over
$GF(q)$ whose parity-check  matrix is given
by~(\ref{pcSD}) and~(\ref{pcSDc}).

\begin{ex}
\label{excode}
{\em
Consider the finite field $GF(16)$ and let $\al$ be a primitive
element, i.e., $\cO(\al)\eq 15$. Then, the parity-check matrix of
$\C^{(0)}(3,5;f_{\al}(x))$ is given by

$$
\left(
\begin{array}{ccccc|ccccc|ccccc}
1&1&1&1&1&0&0&0&0&0&0&0&0&0&0\\
0&0&0&0&0&1&1&1&1&1&0&0&0&0&0\\
0&0&0&0&0&0&0&0&0&0&1&1&1&1&1\\
1&\al&\al^2&\al^3&\al^4&\al^5&\al^6&\al^7&\al^8&\al^9&\al^{10}&\al^{11}&\al^{12}&\al^{13}&\al^{14}\\
1&\al^{14}&\al^{13}&\al^{12}&\al^{11}&
\al^{10}&\al^9&\al^8&\al^7&\al^6&\al^{5}&\al^{4}&\al^{3}&\al^{2}&\al\\
\end{array}
\right)
$$

Similarly, the parity-check matrix of
$\C^{(0)}(5,3;f_{\al}(x))$ is given by

$$
\left(
\begin{array}{ccc|ccc|ccc|ccc|ccc}
1&1&1&0&0&0&0&0&0&0&0&0&0&0&0\\
0&0&0&1&1&1&0&0&0&0&0&0&0&0&0\\
0&0&0&0&0&0&1&1&1&0&0&0&0&0&0\\
0&0&0&0&0&0&0&0&0&1&1&1&0&0&0\\
0&0&0&0&0&0&0&0&0&0&0&0&1&1&1\\
1&\al&\al^2&\al^3&\al^4&\al^5&\al^6&\al^7&\al^8&\al^9&\al^{10}&\al^{11}&\al^{12}&\al^{13}&\al^{14}\\
1&\al^{14}&\al^{13}&\al^{6}&\al^{5}&
\al^{4}&\al^{12}&\al^{11}&\al^{10}&\al^3&\al^{2}&\al &\al^{9}&\al^{8}&\al^7\\
\end{array}
\right)
$$
}
\end{ex}

Let us point out that the construction of this type of codes is valid
also over the ring of polynomials modulo $M_p(x)\eq 1+x+\cdots
+x^{p-1}$, $p$ a prime number, as done with the Blaum-Roth (BR)
codes~\cite{br}. In that case, $\cO(\al)\eq p$, where $\al^{p-1}\eq
1+\al+\cdots +\al^{p-2}$. The construction
proceeds similarly, and we denote it $\C^{(0)}(m,n;M_p(x))$. Utilizing
the ring modulo $M_p(x)$ allows for XOR operations at the encoding
and the decoding without look-up tables in a finite field, which is
advantageous in erasure decoding~\cite{br}. It is well known that
$M_p(x)$ is irreducible if and only if 2 is primitive in
$GF(p)$~\cite{ms}. 

\begin{ex}
\label{exM17}
{\em
Consider the ring of polynomials modulo $M_{17}(x)$ and let $\al$ be an
element in the ring such that $\al^{16}\eq
1+\al+\cdots +\al^{15}$, thus, $\cO(\al)\eq 17$ (notice, $M_{17}(x)$
is reducible). Then, the parity-check matrix of
$\C^{(0)}(4,4;M_{17}(x))$ is given by

$$
\left(
\begin{array}{cccc|cccc|cccc|cccc}
1&1&1&1&0&0&0&0&0&0&0&0&0&0&0&0\\
0&0&0&0&1&1&1&1&0&0&0&0&0&0&0&0\\
0&0&0&0&0&0&0&0&1&1&1&1&0&0&0&0\\
0&0&0&0&0&0&0&0&0&0&0&0&1&1&1&1\\
1&\al&\al^2&\al^3&\al^4&\al^5&\al^6&\al^7&\al^8&\al^9&\al^{10}&\al^{11}&\al^{12}&\al^{13}&\al^{14}&\al^{15}\\
1&\al^{16}&\al^{15}&\al^{14}&\al^{8}&\al^{7}&\al^6&\al^5&\al^{16}&\al^{15}&\al^{14}&\al^{13}&\al^{7}&\al^{6}&\al^5&\al^4\\
\end{array}
\right)
$$
}
\end{ex}

We have the following theorem:

\begin{theo}
\label{theoSD}
{\em
Codes $\C^{(0)}(m,n;f_{\al}(x))$ and $\C^{(0)}(m,n;M_p(x))$ are SD codes.
}
\end{theo}

\pf According to Definition~\ref{defPMDS}, we have to prove first
that 3 erasures in the same row will be corrected. Based on the
parity-check matrix of the code, this will happen if and only if, for
any $0\leq i\leq m-1$ and $0\leq j_0<j_1<j_2\leq n-1$,

\begin{eqnarray*}
\det\left(
\begin{array}{ccc}
1&1&1\\
\al^{in+j_0}&\al^{in+j_1}&\al^{in+j_2}\\
\al^{2in-j_0}&\al^{2in-j_1}&\al^{2in-j_2}\\
\end{array}
\right)&\neq &0
\end{eqnarray*}

But the determinant of this $3\times 3$ matrix can be easily transformed into a Vandermonde
determinant on $\al^{j_0}$, $\al^{j_1}$ and $\al^{j_2}$ times a power of
$\al$, so it is invertible in a field and also in the ring of
polynomials modulo $M_p(x)$~\cite{br}.

Next we have to prove that if we have two erasures in locations $i$
and $j$ of row $\ell$, say, $0\leq i<j\leq n-1$, and two erasures
in locations $i'$ and $j'$ of row $\ell'$, $0\leq i'<j'\leq n-1$,
$0\leq \ell<\ell'\leq m-1$, such that, either $i\eq i'$, $i\eq
j'$, $j'\eq i$ or $j\eq j'$, then

\begin{eqnarray*}
\det\left(
\begin{array}{cccc}
1&1&0&0\\
0&0&1&1\\
\al^{\ell n+i}&\al^{\ell n+j}&\al^{\ell' n+i'}&\al^{\ell' n+j'}\\
\al^{2\ell n-i}&\al^{2\ell n-j}&\al^{2\ell' n-i'}&\al^{2\ell' n-j'}\\
\end{array}
\right)&\neq &0
\end{eqnarray*}

After some row manipulation, the inequality above holds if and only
if

\begin{eqnarray*}
\det\left(
\begin{array}{cc}
\al^{\ell n+i}\left(1\xor\al^{j-i}\right)&\al^{\ell' n+i'}\left(1\xor\al^{j'-i'}\right)\\
\al^{2\ell n-j}\left(1\xor\al^{j-i}\right)&\al^{2\ell' n-j'}\left(1\xor\al^{j'-i'}\right)\\
\end{array}
\right)&\neq &0.
\end{eqnarray*}
$1\xor\al^{j-i}$ is invertible in $GF(q)$ since $1\leq j-i<\cO(\al)$,
but the same is true in the
polynomials modulo $M_p(x)$~\cite{br}, thus, the inequality above is
satisfied if and only if

\begin{eqnarray*}
\det\left(
\begin{array}{cc}
\al^{i}&\al^{i'}\\
\al^{-j}&\al^{(\ell'-\ell) n-j'}\\
\end{array}
\right)&\neq &0.
\end{eqnarray*}

Assume that this determinant is 0. Redefining $\ell\,\la\,\ell'-\ell$, then
$1\leq\ell\leq m-1$ and we have

\begin{eqnarray*}
\al^{\ell n}&=&\al^{i'+j'-i-j}.\\
\end{eqnarray*}

We will show that this is not possible. Assume that $i\eq i'$. Then,

\begin{eqnarray*}
\al^{\ell n}&=&\al^{j'-j}.\\
\end{eqnarray*}

Assume that $j'\geq j$. Then, $\ell n\eq j'-j$, a contradiction since
$j'-j\leq n-1$ and $n\leq \ell n<mn\leq \cO(\al)$.

So, assume $j'<j$. Then, $\ell n\eq \cO(\al)+j'-j$. But this also gives a
contradiction, since $\ell n\leq mn-n\leq \cO(\al)-n$, and
$\cO(\al)+j'-j\geq \cO(\al)-n+1$.

The cases $i\eq j'$, $j'\eq i$ and $j\eq j'$ are handled similarly.
\qed

Next we show how to construct PMDS codes.

Let $\al\in GF(2^b)$ and $2mn\leq \cO(\al)$. Consider the $(m+2)\times
mn$ parity-check matrix given by~(\ref{pcSD}) and, for $0\leq i\leq m-1$,

\begin{eqnarray}
\label{pcPMDS}
\uc_{in},\uc_{in+1},\ldots ,\uc_{(i+1)n-1}&=&\left(
\begin{array}{cccccc}
\ue_i &\ue_i &\ldots &\ue_i & \ldots &\ue_i \\
\al^{2in}&\al^{2in+1}&\ldots &\al^{2in+j}&\ldots &\al^{2(i+1)n-1}\\
\al^{4in}&\al^{4in-1}&\ldots &\al^{4in-j}&\ldots &\al^{(4i-1)n+1}\\
\end{array}
\right)
\end{eqnarray}

We denote the $[mn,m(n-1)-2]$ code over $GF(q)$ whose parity-check
matrix is given
by~(\ref{pcSD}) and~(\ref{pcPMDS}) as $\C^{(1)}(m,n;f_{\al}(x))$. The same
can be done with the ring of polynomials modulo $M_p(x)$, in which
case we denote the code $\C^{(1)}(m,n;M_p(x))$.

\begin{ex}
\label{exM172}
{\em
As in Example~\ref{exM17}, consider the ring of polynomials modulo
$M_{17}(x)$ and let $\al$ be an
element in the ring such that $\cO(\al)\eq 17$ and $\al^{16}\eq
1+\al+\cdots +\al^{15}$. Then, the parity-check matrix of
$\C^{(1)}(2,4;M_{17}(x))$ is given by

$$
\left(
\begin{array}{cccc|cccc}
1&1&1&1&0&0&0&0\\
0&0&0&0&1&1&1&1\\
1&\al&\al^2&\al^3&\al^8&\al^9&\al^{10}&\al^{11}\\
1&\al^{15}&\al^{14}&\al^{13}&\al^{16}&\al^{15}&\al^{14}&\al^{13}\\
\end{array}
\right)
$$
}
\end{ex}

\begin{theo}
\label{theoPMDS}
{\em
Codes $\C^{(1)}(m,n;\al;q)$ and $\C^{(1)}(m,n;M_p(x))$ are PMDS codes.
}
\end{theo}

\pf As in Theorem~\ref{theoSD}, we have to prove first
that three erasures in the same row will always be corrected.

Based on the
parity-check matrix of the code, this will happen if and only if, for
any $0\leq i\leq m-1$ and $0\leq j_0<j_1<j_2\leq n-1$,

\begin{eqnarray*}
\det\left(
\begin{array}{ccc}
1&1&1\\
\al^{2in+j_0}&\al^{2in+j_1}&\al^{2in+j_2}\\
\al^{4in-j_0}&\al^{4in-j_1}&\al^{4in-j_2}\\
\end{array}
\right)&\neq &0
\end{eqnarray*}

Again, the determinant of this $3\times 3$ matrix can be transformed into a Vandermonde
determinant on $\al^{j_0}$, $\al^{j_1}$ and $\al^{j_2}$ times a power of
$\al$, so it is invertible in a field and also in the ring of
polynomials modulo $M_p(x)$.

Next we have to prove that if we have two erasures in locations $i$
and $j$ of row $\ell$, say, $0\leq i<j\leq n-1$, and two erasures
in locations $i'$ and $j'$ of row $\ell'$, $0\leq i'<j'\leq n-1$,
$0\leq \ell<\ell'\leq m-1$, then

\begin{eqnarray*}
\det\left(
\begin{array}{cccc}
1&1&0&0\\
0&0&1&1\\
\al^{2\ell n+i}&\al^{2\ell n+j}&\al^{2\ell' n+i'}&\al^{2\ell' n+j'}\\
\al^{4\ell n-i}&\al^{4\ell n-j}&\al^{4\ell' n-i'}&\al^{4\ell' n-j'}\\
\end{array}
\right)&\neq &0
\end{eqnarray*}

After some row manipulation, the inequality above holds if and only
if

\begin{eqnarray*}
\det\left(
\begin{array}{cc}
\al^{2\ell n+i}\left(1\xor\al^{j-i}\right)&\al^{2\ell' n+i'}\left(1\xor\al^{j'-i'}\right)\\
\al^{4\ell n-j}\left(1\xor\al^{j-i}\right)&\al^{4\ell' n-j'}\left(1\xor\al^{j'-i'}\right)\\
\end{array}
\right)&\neq &0
\end{eqnarray*}
Again, $1\xor\al^{j-i}$ is invertible in $GF(q)$ and
in the ring of polynomials modulo $M_p(x)$, thus, the
inequality above is satisfied if and only if

\begin{eqnarray*}
\det\left(
\begin{array}{cc}
\al^{i}&\al^{i'}\\
\al^{-j}&\al^{2(\ell'-\ell) n-j'}\\
\end{array}
\right)&\neq &0.
\end{eqnarray*}

Assume that this determinant is 0. Redefining $\ell\,\la\,\ell'-\ell$, then
$1\leq\ell\leq m-1$ and we have

\begin{eqnarray*}
\al^{2\ell n}&=&\al^{i'+j'-i-j}.\\
\end{eqnarray*}

But this is not possible. In effect, assume first that $i'+j'\geq
i+j$. Then, since $2n\leq 2\ell n< \cO(\al)$, we would have
$2\ell n\eq i'+j'-i-j$, a contradiction since
$i'+j'-i-j\leq 2(n-1)$.

So, assume $i'+j'<i+j$. Then, $2\ell n\eq
\cO(\al)+i'+j'-i-j$. This also gives a
contradiction, since $2\ell n\leq 2mn-2n\leq \cO(\al)-2n$, and
$\cO(\al)+i'+j'-i-j\geq \cO(\al)-2n+2$. \qed

\section{Conclusions}
We have presented constructions of PMDS and SD codes extending RAID 5
with two extra parities, solving an open problem since previous
constructions were based on computer search. It is an open problem to
extend the results to more parities.

\end{document}